\documentclass[useAMS,usenatbib]{mn2e}
\usepackage{graphicx}
\usepackage{dcolumn}
\usepackage{bm}

\usepackage{amsmath}
\usepackage{amssymb}

\title[Velocity distribution of dark matter particles]{Principal properties of the velocity distribution of dark matter particles on the outskirts of
the Solar System}

\author[A. N. Baushev]{A. N. Baushev\\
DESY, 15738 Zeuthen, Germany\\
 Institut f\"ur Physik und Astronomie, Universit\"at Potsdam, 14476
Potsdam-Golm, Germany\\}
\begin{document}

\date{}

\pagerange{\pageref{firstpage}--\pageref{lastpage}}
\pubyear{2011}

\maketitle

\label{firstpage}

\begin{abstract}
The velocity distribution of the dark matter particles on the outskirts of the Solar System remains
unclear. We suggest to determine it using experimentally found properties of the oldest halo
objects. Indeed, the oldest halo stars and globular clusters form a collisionless system, as well
as dark matter particles do, and they evolved in the same gravitational field. If we accept this
analogy, we can show that the velocity distribution of the dark matter particles should be highly
anisotropic and have a sharp maximum near $\upsilon\sim 500$~{km/s}. The distribution is totally
different from the Maxwell one.

We analyze the influence of the distribution function on the results of dark matter detection
experiments. It is found that the direct detection signal should differ noticeably from the one
calculated from the Maxwell distribution with $\langle \upsilon\rangle \simeq 220$~{km/s}, which is
conventional for direct detection experiments (the ratio depends on the detector properties and
typically falls within the range between $6$ and $0.2$). Moreover, the sharp distinction from the
Maxwell distribution can be very essential to the observations of dark matter annihilation.
\end{abstract}

\begin{keywords}
cosmology: dark matter, elementary particles, methods: analytical.
\end{keywords}

\section{Introduction}
One of the most evident manifestations of the dark matter existence is the detection of huge
invisible halos (with density profiles $\rho\sim r^{-2}$) surrounding galaxies \citep{suchkov}. Our
Galaxy also has such a halo. We symbolize the orbital radius of the Solar System, the average
velocity of the Galaxy rotation, and the escape velocity at this radius by  $r_{\odot}$,
$\upsilon_{\odot}$, and $\upsilon_{esc}$, respectively. We also denote the radial and tangential
components of a dark matter particle velocity by $\upsilon_r$ and
$\upsilon_\rho\equiv\sqrt{\upsilon^2_\phi+\upsilon^2_\theta}$. Velocity distribution of the dark
matter particles (hereafter DMPs) inside the halo is poorly known; it is usually supposed to be
Maxwell with a cut-off when $\upsilon>\upsilon_{esc}$ \citep{direkt}.
\begin{equation}
f(\upsilon) = \frac{N}{(\sqrt{\pi} \upsilon_{\odot})^3} \exp \left(-
\frac{\upsilon^2}{\upsilon_{\odot}^2}\right), \quad \upsilon<\upsilon_{esc}
 \label{11b1}
\end{equation}
We accept $r_{\odot}=8$~{kpc}, $\upsilon_{\odot}=220$~{km/s}, $\upsilon_{esc}=643$~{km/s}. $N$ is a
normalizing constant, and for the chosen parameters $N\simeq 1.001$. It is worthy of noting that
$\upsilon_{orb}$ remains almost constant throughout the halo.

Distribution (\ref{11b1}) faces with difficulties. In fact, in the framework of collisionless
dynamics it can be naturally obtained from profile $\rho\sim r^{-2}$, subject to the condition,
however, that function $f$ is isotropic, i.e. $f$ depends only on $|\upsilon|$ (so called
isothermal model). This assumption seems highly improbable. Indeed, if (\ref{11b1}) is true, the
majority of the dark matter particles has large specific angular momentum $\mu\equiv
\mathfrak{M}/m_{\chi} = [\upsilon\times r]$. The average angular momentum of the particles (and,
consequently, of the halo) is zero. However, the root-mean-square momentum is $\sqrt{\langle
{\mu}^2\rangle }\simeq 1800\; \text{kpc}\cdot\text{km/s}$ at $r_{\odot}$. Moreover, since in this
model $\mu\sim r \upsilon_{orb}$ and $\upsilon_{orb}$ is constant in the halo if $\rho\sim r^{-2}$,
the root-mean-square angular momentum reaches an incredibly huge value $\sqrt{\langle
{\mu}^2\rangle }\sim 4\cdot 10^4\; \text{kpc}\cdot\text{km/s}$ at the edge of the halo ($r\sim
200$~{kpc}). Meanwhile, according to modern cosmological conceptions, not only the total angular
momentum of the halo but also the momentum of each particle should have been negligibly small on
the linear stage of the structure formation \citep{gorbrub2}. The halo could gain some angular
momentum later, as a result of tidal perturbations or merging of smaller halos; numerical
simulations show, however, that it cannot be large \citep{maccio}.

Some results of stellar dynamics are frequently used in order to show that the dark matter
particles could gain a large angular momentum during the Galaxy evolution. The parallels between
DMP and stellar dynamics, however, are not universally true. The point is that stars are compact
objects, and their gravitational field can be strong, at least, locally. On the contrary, the
small-scale gravitational field of the dark matter is always small \citep{2009}. Therefore
important relaxation mechanisms of stellar systems like close pair approaches or an interaction
with the interstellar medium are completely ineffective for DMPs. So called violent relaxation
\citep{violent} is perhaps the only way to impart significant angular momentum to the dark matter
particles. However, it acts also on the halo stars; moreover, its efficiency decreases with radius,
and it should stronger affect the star distribution since the stellar halo is more compact.

To summarize: all halo objects initially had $\upsilon_\rho\simeq 0$ and later gained some angular
momentum because of various processes like relaxation or tidal effects. All the mechanisms
increased $\langle \upsilon^2_\rho \rangle $ of the halo stars, at least, as much as of the DMPs,
while some mechanisms affected only the stars, and not the DMPs. Consequently, velocity
distribution of the oldest stellar halo population should be closer to the Maxwellian one, than the
distribution of the dark matter particles. In particular, the tangential velocity dispersion
$\sigma(\upsilon_\rho)$ of the DMPs cannot be larger than that of the halo stars at the same
radius.

Modern observations of the oldest halo stars -- subdwarfs -- confirm the above reasoning
\citep{2009MNRAS.399.1223S}. Their tangential dispersion
$\sigma(\upsilon_\rho)\equiv\sigma_0\simeq 80$~{km/s}, which corresponds to $\sqrt{\langle
{\mu}^2\rangle }\simeq 900\; \text{kpc}\cdot\text{km/s}$, is two times lower than in (\ref{11b1}).
Moreover, the distribution widely differs from the Maxwellian: $\sigma(\upsilon_r)$ is much larger
than $\sigma(\upsilon_\rho)$. Consequently, $\sigma(\upsilon_\rho)$ of the dark matter particles on
the outskirts of the Solar System does not exceed $\sigma_0=80$~{km/s} and can be even smaller.
Second, the observations show that the halo stars have not yet relaxed and their orbits are rather
prolate. On the other hand, if distribution (\ref{11b1}) is correct, the ellipticity of the
majority of DMP orbits is small, and then the dark matter is the only class of halo objects that
move almost circularly. It seems much more natural to assume the opposite, and we come to the
premises we use throughout this article:

1) The specific angular momentum $\mu$ of, at least, the main part of the particles is fairly
small, and their orbits are rather prolate (below we will express this supposition quantitatively).

2) The Galactic halo is stationary and spherically symmetrical. The latter supposition is not quite
accurate: a part of dark matter can form a so-called thick disk \citep{thickdisk}, moreover, the
influence of the star disk also takes place. However, our assumption is quite acceptable for an
estimative consideration.

3) In accordance with observations \citep{2010MNRAS.407....2D},  we assume that the density profile
of the Galaxy is $\rho\propto r^{-2}$ up to some large enough radius $R$. The profile cannot be
valid for an arbitrary large $r$, since in the opposite case the halo mass would be infinite.
Starting from some radius (we denote it by $R$), the halo density drops much faster. As we will
see, this is the main parameter defining the DMP velocity distribution. The profile $\rho\sim
r^{-2}$ has an another demerit: the predicted annihilation signal (which is proportional to
$\rho^2$) diverges when $r\to 0$. However, as it will be shown below, the velocity distribution of
DMPs on the outskirts of the Solar System weakly depends on the density profile inside the solar
orbit. Therefore we will not discuss a very complex question of the dark matter profile near the
galactic centre.

4) We assume that the dark matter mass outside of the radius $R$ is negligibly small as compared
with the total halo mass. This supposition is the most discussable; however, there are strong
arguments in favour of it. First, experimental data show \citep{binneytremaine} that $\rho\sim
r^{-2}$ up to very large distances. Second, in order that the total halo mass is finite, the
density must fall much faster than $r^{-2}$ (at least, faster than $r^{-3}$) at large distances.
For instance, one of the most popular Einasto profile predicts an exponential density decreasing
\citep{einasto}. Third, the outer regions of the halo are respectively weakly bound in the
gravitational field and can easily be torn away by tidal effects.

These four suppositions turn out to be sufficient to derive the velocity distribution more or less
unambiguously. All our calculations are, of course, focused on determination of the DMP velocity
distribution on the outskirts of the Solar System.
\begin{figure}
 \resizebox{\hsize}{!}{\includegraphics[angle=0]{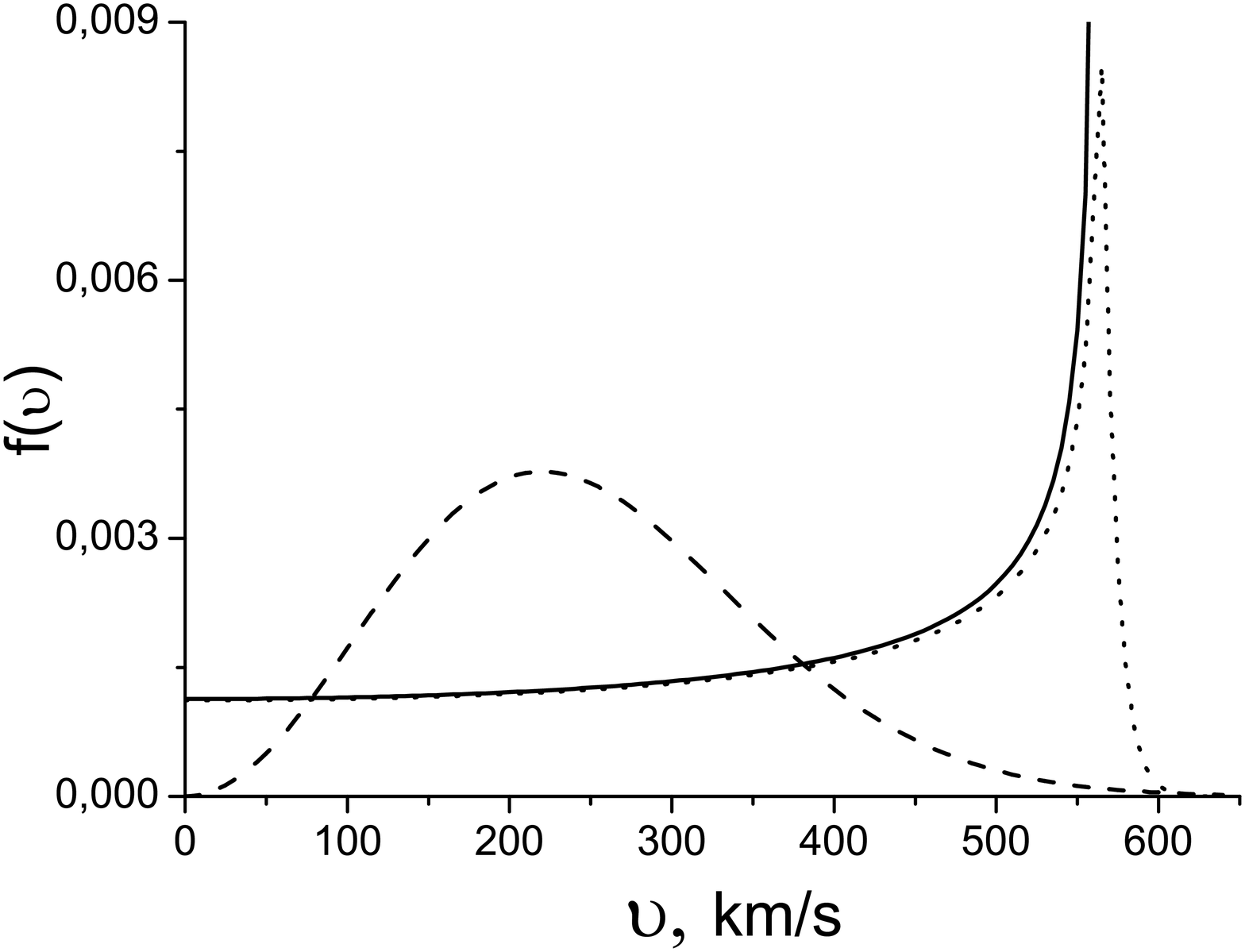}}
 \caption{Anisotropic (\ref{11b2}) (solid line) and Maxwell ((\ref{11b1}), dashed line) distributions of dark matter
 particles over the velocity absolute value. The dotted line represents the velocity distribution if the halo edge
 is smoothed according to (\ref{11g1}).}
 \label{fig1}
\end{figure}
\section{Calculations}
For our Galaxy we accept $R=210$~{kpc}, which corresponds to the total mass of the Galaxy
$M=2.3\cdot 10^{12}M_\odot$. The module of gravitational potential on the edge of the halo is equal
$\Phi=GM/R$. It is easy to see that $\Phi=\upsilon^2_{\odot}$. Since $\rho\sim r^{-2}$, the mass
inside some radius can be found as $\frac{r}{R}M$, gravitational field is equal $\vec
g=G\frac{M}{rR}$, and we obtain the gravitational potential inside the halo:
\begin{equation}
\phi=-\Phi \left(1+\ln\frac{R}{r}\right)
 \label{11a2}
\end{equation}
Let us start our consideration from the case when the DMPs have no angular momentum at all. Then
their trajectories are radial ($\upsilon=|\upsilon_{r}|$), and the task becomes one-dimensional.
Therefore the particle distribution in the halo can be entirely described by a single function
$\psi(r,\upsilon)$, so that $\psi(r,\upsilon) dr d\upsilon$ gives the total mass of dark matter in
the element of phase space $dr d\upsilon$. $\psi$ differs from the standard distribution function
only by a insignificant multiplier --- the DMP mass $m_\chi$. Since we suppose that the dark matter
particles are propelled only by the gravitational force, $m_\chi$ is not important for us, the
calculations are formally valid even for the dark matter consisting of primordial black holes, and
such a definition of $\psi$ allows us to avoid of the undesirable dependence on the DMP mass. The
halo density is bound with the function $\psi$ by a trivial relation
\begin{equation}
\eta\equiv 4\pi r^2 \rho(r)=\int_{0}^\infty \psi(r,\upsilon) d\upsilon
 \label{11a3}
\end{equation}
Here we introduced a more convenient variable $\eta$ instead of $\rho$, $\eta=\it{const}$ if
$\rho\sim r^{-2}$. In our case $\eta=M/R$, and we obtain a determining condition on $\psi$:
\begin{equation}
\int_{0}^\infty \psi(r,\upsilon) d\upsilon = \frac{M}{R}
 \label{11a4}
\end{equation}
Each dark matter particle executes a radial oscillation around the galactic centre. We denote by
$r_0$ the maximum distance it moves off the centre. Its velocity in potential (\ref{11a2}) is equal
to
\begin{equation}
|\upsilon_{r}|=\upsilon=\sqrt{2\Phi\ln\frac{r_0}{r}}
 \label{11a5}
\end{equation}
and we obtain useful equations:
\begin{equation}
r_0=r\exp\left(\frac{\upsilon^2}{2\Phi}\right),\quad \frac{\partial r_0}{\partial
\upsilon}=\frac{\upsilon r_0}{\Phi}, \quad T=r_0 \sqrt\frac{\pi}{2\Phi} \label{11a7}
\end{equation}
Here $T$ is the time required for the particle to fall from $r_0$ to the centre. We introduce a
distribution function $\xi$ of the particles throughout parameter $r_0$, so that $\xi(r_0) dr_0$ is
the total mass of DMPs which apoapsis lies in the interval $[r_0;r_0+dr_0]$. The r-coordinate of
these particles varies between $0$ and $r_0$, and they give a yield into the halo density over all
this interval. Indeed, the fraction of time the DMPs from the subsystem under consideration pass in
an interval $[r;r+dr]$ is equal to $dt/T=dr/(\upsilon T)$. Since the total particle mass of the
subsystem is $\xi(r_0) dr_0$, the contribution to the halo mass in interval $dr$ is equal to
\begin{equation}
dM=\frac{\xi(r_0)}{\upsilon T} dr dr_0
 \label{11a8}
\end{equation}
Velocity interval $d\upsilon$ that is covered by the particles of the subsystem at radius $r$ is
$\frac{\partial r_0}{\partial \upsilon} d\upsilon = dr_0$. Substituting this to (\ref{11a8}) and
taking into account that $dM=\psi(r,\upsilon) dr d\upsilon$, we obtain the general equation for
$\psi(r,\upsilon)$:
\begin{equation}
\psi(r,\upsilon)=\dfrac{\xi(r_0)\frac{\partial r_0}{\partial \upsilon}}{\upsilon T}
 \label{11a9}
\end{equation}
Now we substitute here equations (\ref{11a5}) and (\ref{11a7}):
\begin{equation}
\psi(r,\upsilon)=\sqrt\frac{2}{\pi \Phi} \xi\left[r\exp\left(\frac{\upsilon^2}{2\Phi}\right)\right]
 \label{11a10}
\end{equation}
We can easily find function $\xi$ if we suppose that the halo boundary is sharp, i.e. the density
obeys the law $\rho\sim r^{-2}$ up to radius $R$ and is equal to zero just after it. Then the
maximum velocity the halo particles may have at radius $r$ is
\begin{equation}
\upsilon_{max}=\sqrt{2\Phi\ln\frac{R}{r}}=\upsilon_{\odot} \sqrt{2\ln\frac{R}{r}}
 \label{11a11}
\end{equation}
It is a matter of direct verification to prove that function
\begin{equation}
\xi(r_0)= \frac{M}{\sqrt{\pi}R\sqrt{\ln\frac{R}{r_0}}}
 \label{11a6}
\end{equation}
satisfies condition (\ref{11a4}). This function has a peculiarity at $r_0=R$. So the main part of
DMPs comes to us from the very edge of the halo. The distribution function is
\begin{equation}
\psi(r,\upsilon)= \frac{2 M}{\pi R\sqrt{\upsilon^2_{max}-\upsilon^2}}
 \label{11a12}
\end{equation}
where $\upsilon_{max}$ is defined by (\ref{11a11}), $\upsilon\in [0;\upsilon_{max}]$. The
distribution through the radial velocity $\upsilon_{r}$ can be easily obtained from (\ref{11a12}):
$\psi(r,\upsilon_{r})=\psi(r,-\upsilon_{r})=\psi(r,\upsilon)/2$, where $\upsilon_r\in
[-\upsilon_{max};\upsilon_{max}]$. Normalized velocity distribution on the outskirts of the Solar
System is equal to
\begin{equation}
f(\upsilon)=
\frac{2}{\pi\sqrt{\upsilon^2_{max}-\upsilon^2}}\simeq\frac{2}{\pi\sqrt{(2.2\upsilon_{\odot})^2-\upsilon^2}}
 \label{11b2}
 \end{equation}
The distribution has a peculiarity at $\upsilon_{max}\simeq 2.55\upsilon_{\odot}\simeq 562$~{km/s},
which is a result of the supposition that the density after $r=R$ immediately drops to zero. In
actuality there is a characteristic length $l$ of density decreasing out of $R$, for instance, if
the decreasing is exponential $\rho\propto exp(-r/r_d)$ we can use $r_d$ as $l$. Then the cusp at
$\upsilon_{max}$ transforms into a smooth peak. Since the free fall acceleration at $r=R$ is
$g=GM/R^2$, we can easily estimate the width of the peak:
\begin{equation}
\frac{\delta\upsilon}{\upsilon}\simeq\frac12\frac{gl}{\upsilon^2}=\frac{l}{4R\ln\frac{R}{r}}
 \label{11b7}
 \end{equation}
Near the Solar System $\delta\upsilon\simeq (l/R)\cdot 50$~{km/s}. It is natural to assume that
$l\ll R$ (actually, it follows from supposition 4 in the {\it Introduction}), so the distribution
is still very narrow.

To illustrate the above reasoning, let us consider the following density profile:
\begin{equation}
dM/dr=\begin{cases}
{\it const},&\text{if $r\le R$;}\\
{\it const}\cdot\exp\left(-\dfrac{r-R}{r_d}\right), &\text{if $r>R$}
\end{cases}
\label{11g1}
\end{equation}
In this case, we cannot determine the distribution $\xi(r)$ analytically. However, if $r_d\ll R$,
we can neglect the change in the gravitational field on the scale of $r_d$ and then find an
approximate solution. Fig.~\ref{fig1} represents distributions (\ref{11b2}) (for the sharp halo
edge, solid line) and for density profile (\ref{11g1}) (dotted line, we accepted $r_d=0.1 R=
21$~{kpc}). One can see that in the case of smoothed halo edge the peculiarity at
$\upsilon_{max}\simeq 562$~{km/s} is transformed into a smooth peak, which is, however, quite
narrow, and the general shape of the distribution remains the same. We will discuss the influence
of the halo density profile on the velocity distribution in the beginning of the {\it Discussion}
section.

\begin{figure}
 \resizebox{\hsize}{!}{\includegraphics[angle=0]{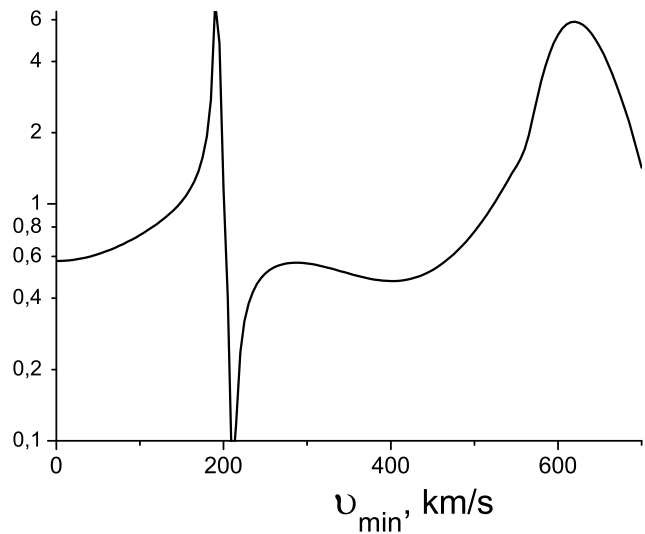}}
 \caption{The ratio between the double amplitudes $2A=I_{max}-I_{min}$ of direct detection signals (\ref{11b8}) calculated for
 anisotropic distribution (\ref{11b6}) and Maxwell distribution (\ref{11b1}).}
 \label{fig2}
\end{figure}

 The angular momentum of the dark
matter particles hardly can be exactly equal to zero. If a particle possesses some specific
momentum $\mu$, its velocity in gravitational field (\ref{11a2}) is equal to:
\begin{equation}
\upsilon_\rho=\frac{\mu}{r};\quad |\upsilon_{r}|=
\sqrt{2\Phi\ln\frac{r_0}{r}-\mu^2\left(\frac{1}{r^2}-\frac{1}{r_0^2}\right)} \label{11b3}
\end{equation}
Since (\ref{11b3}) is distinct from (\ref{11a5}), distribution (\ref{11a6}), strictly speaking, is
not valid anymore. However, we may use it if the difference between (\ref{11b3}) and (\ref{11a5})
is small. Mathematically it can be written as:
\begin{equation}
2\Phi\ln\frac{r_0}{r}\gg \mu^2\left(\frac{1}{r^2}-\frac{1}{r_0^2}\right)
 \label{11b5}
 \end{equation}
As we have already discussed in the {\it Introduction}, specific angular momentum of the majority
of the dark matter particles hardly can be larger than $900\;\text{kpc}\cdot\text{km/s}$.
Substituting $r_0=R$, $r=r_{\odot}$ to the inequality, we can see that its right part is equal to
$110^2\; \text{km}^2/\text{s}^2$, while its left part is $\sim 550^2\; \text{km}^2/\text{s}^2$.
Hence the inequality asserts near the Solar System, the influence of the angular momentum on the
radial dynamics is still negligible for the majority of DMPs, and we can use (\ref{11a6}) up to
$r_\odot$ as before. Thus the DMP distribution throughout $\upsilon_r$ in this approximation
coincides with (\ref{11b2}). However, the particles have also some distribution throughout
$\upsilon_\rho$. For simplicity we will suppose that $f\propto \exp(-\upsilon^2_\rho/2\sigma^2_0)$
where $\sigma_0=80$~{km/s}, though the distribution can be much narrower. Then the normalized DMP
distribution near the Solar System can be closely approximated by
\begin{equation}
f(\upsilon)=
\frac{\exp\left(-\dfrac{\upsilon^2_\rho}{2\sigma^2_0}\right)}{2\pi^2\sigma^2_0\sqrt{\upsilon^2_{max}-\upsilon_r^2}}
 \label{11b6}
 \end{equation}
where $\upsilon_r\in [-\upsilon_{max};\upsilon_{max}]$, $\upsilon_{max}=562$~{km/s}. Distribution
(\ref{11b6}) is strongly anisotropic and actually describes two colliding beams of particles.

\section{Discussion}
Fig.~\ref{fig1} represents distributions (\ref{11b2}) and (\ref{11b1}) (solid and dashed lines,
respectively). One can see that (\ref{11b2}) is much narrower and has much higher average velocity.
The physical reason of it is obvious: in the case of Maxwell distribution (\ref{11b1}) the
particles move almost circularly, which is why only a few of DMPs from the edge of the halo reach
the Solar orbit. On the contrary, in the case considered in this article the majority of DMPs comes
from the halo edge and thus are much more accelerated by the gravitational field. Consequently, the
question of what of the distributions, (\ref{11b1}) or (\ref{11b2}), is correct, can be reduced to
whether the particles from the halo edge can reach the Solar orbit or not. In addition to the
arguments presented in the {\it Introduction} we note that, according to (\ref{11b3}), a particle
falling from $r=R$ should have a specific angular momentum $\mu\sim 4000\;
\text{kpc}\cdot\text{km/s}$, lest the particle can reach $r=8$~{kpc}. This value is huge, it far
exceed not only the characteristic momentum of halo objects, but even the momentum of the disk, and
thus looks very unlikely. So particles from the edge of the halo freely reach the Earth, and their
spectrum should be closer to (\ref{11b2}).

A similar consideration allows us to examine the dependence of velocity distribution (\ref{11b2})
on the density profile. Our assumption of the existence of a large region with $\rho\propto r^{-2}$
is approximately correct for massive spiral galaxies, such as the Milky-Way
\citep{2010MNRAS.407....2D}. However, we obtained (\ref{11b2}) on the additional assumption that
the edge of the halo is more or less sharp. Meanwhile, the outer region of the halo can have a
density profile steeper than $r^{-2}$, but not steep enough to be considered as a cutoff. As an
instance, one can consider a double power-law halo \citep{lisanti}. How can it influence on the
velocity profile? The answer depends on the mass fraction of this steeper region with respect to
the total mass of the halo. If the fraction is small, the distribution differs little from
(\ref{11b2}), as we demonstrated with distribution (\ref{11g1}). Let us consider the case when the
fraction is significant. We indicate the radius where the profile gets steeper than $r^{-2}$ by
$\mathfrak R$; hereafter we will name 'outer halo' the region out of  $\mathfrak R$. As we could
see, in the model with sharp halo edge the majority of the particles comes from the edge of the
halo. Expressing this fact mathematically, distribution $\xi(r_0)$ is small for $r_0<R$ and goes to
infinity at the edge of the halo (\ref{11a6}). It is easy to show that in the case of the presence
of a massive outer halo dark matter particles mainly come from it, and the fraction of the
particles with $r_0<\mathfrak R$ is respectively small. Let us consider a system of particles with
$r_0>\mathfrak R$. According to (\ref{11a8}), their contribution to the halo mass in interval $dr$
depends on $r$ only as $\upsilon_r^{-1}(r)$. $\upsilon_r^{-1}(r)$, however, changes rather slowly
inside the region where $\rho\propto r^{-2}$, since the potential there (see (\ref{11a2})) depends
on $r$ only logarithmically. Therefore, the particles falling from the edge of halo provide almost
the same contribution to the halo mass on each radius inside the region $dM\approx \text{\it
const}$, which corresponds to $\rho\approx r^{-2}$. Function $\xi(r_0)$ should be chosen so that it
reproduces the density profile, in particular, it should provide $\rho\propto r^{-2}$. However, as
we could see, the particles from the outer halo by themselves give a very similar profile, and we
need relatively few of particles with $r_0<\mathfrak R$ in order to make it exactly $r^{-2}$.
Consequently, $\xi(r_0)$ is small for $r_0<\mathfrak R$, and a significant fraction still comes
from the halo edge. Thus, this property of the distribution does not depend on the exact density
profile, being only a result of the assumption of strong anisotropy of the velocity distribution
and of the flatness of potential (\ref{11a2}).

Function $\xi(r_0)$ unambiguously determines the velocity distribution of the dark matter
particles, and the above-mentioned common properties of $\xi(r_0)$ directly correspond to
characteristics of $f(\upsilon)$. For the Milky Way galaxy $\mathfrak R$ is large ($\mathfrak R >
100$~{kpc}, \citet{2010MNRAS.407....2D}), which corresponds to $\upsilon\simeq 500$~{km/s} for a
terrestrial observer. Below this speed the distribution as a whole is similar to (\ref{11b2}), and
 it does not depend much on the density profile of the outer halo, since it is created by the
 particles with $r_0<\mathfrak R$. On the contrary, the distribution in region $\upsilon_{esc}>\upsilon>
 500$~{km/s} strongly depends on the density distribution on the edge of the halo and can differ
drastically from (\ref{11b2}). However, now we can estimate what distribution $f(\upsilon)$ looks
like if profile $\rho\propto r^{-2}$ is valid only up to $\mathfrak R \simeq 100$~{kpc}, and then
the halo has a massive outer region, for instance, a Navarro-Frank-White tail. We can expect that
$f(\upsilon)$ is similar to (\ref{11b2}) below $\upsilon= 500$~{km/s} and is totaly defined by the
density profile of the outer halo for $\upsilon_{esc}>\upsilon> 500$~{km/s}. Since the outer halo
is rather extensive, we can expect that a cusp in (\ref{11b2}) is strongly smoothed,
Fig.~\ref{fig1} illustrates all these properties, though we made rather a small modification of the
density profile (\ref{11g1}). So the main characteristic features of distribution (\ref{11b2}) are
not sensitive to the density profile: the distribution is completely not Maxwell, a significant
fraction of particles comes from the edge of the halo forming a high-velocity bump in
$f(\upsilon)$.

The difference between (\ref{11b1}) and (\ref{11b2}) is important for various aspects of the dark
matter physics. In the case of direct dark matter search the signal, roughly speaking, can be
represented as a product of a part depending almost not at all on the DMP distribution and an
integral \citep{Belanger}
\begin{equation}
I(\upsilon)=\int^\infty_{\upsilon_{min}}\frac{\tilde f(\upsilon)}{\upsilon}\; d\vec\upsilon
 \label{11b8}
 \end{equation}
Here $\upsilon_{min}$ is the minimal DMP speed, to which the detector is sensitive, $\tilde
f(\upsilon)$ is the distribution in the Earth's frame of reference, obtained from (\ref{11b1}) or
(\ref{11b2}) by a Galilean transformation (see details in \citet{direkt}, section 3.3). Because of
the Earth's orbital motion $I$ varies with a year period, and it is this variation that is observed
in the direct detection experiments. Fig.~\ref{fig2} shows the ratio between the double amplitudes
$2A=I_{max}-I_{min}$ of direct detection signals calculated for anisotropic distribution
(\ref{11b6}) and Maxwell distribution (\ref{11b1}), as a function of $\upsilon_{min}$. One can see
a very significant difference.

The difference between distributions (\ref{11b1}) and (\ref{11b2}) can also be important for the
indirect dark matter search. Neutrino observations, for instance, are trying to detect the dark
matter annihilation going on in the centre of the Sun. The signal depends on the number of the DMPs
captured by the Sun. Very roughly speaking, it is proportional to $\tilde f(0)$. One can see that
for distribution (\ref{11b2}) $\tilde f(0)$ is approximately $20$ times smaller than for the
conventional Maxwell distribution (\ref{11b1}).

If the $s$-channel of the dark matter annihilation dominates (which is typical), $\langle \sigma
\upsilon\rangle \simeq{\it const}$, and the signal is not sensitive to DMP distribution. However,
if the $p$-channel prevails, $\sigma\simeq{\it const}$, and the signal is proportional to the
averaged velocity of the particle collision. $\langle v_c\rangle \simeq 0.8\upsilon_{max}$ in the
case of distribution (\ref{11b2}) and $\langle v_c\rangle =\sqrt{8/\pi}\; \upsilon_{orb}$ in the
case of Maxwell distribution (\ref{11b1}). $\upsilon_{max}$ relates to $\upsilon_{orb}$ by
(\ref{11a11}), and one can see that $\langle v_c\rangle$ predicted by anisotropic distribution
(\ref{11b2}) is much lower at the halo edge and much higher in the central region than the Maxwell
one. Near the Solar System, however, they are nearly equal. Finally, Sommerfeld effect is inversely
proportional to $\langle v_c\rangle$, and if it plays any role in the dark matter annihilation, it
is also sensitive to the particle distribution.

\section{Acknowledgements}
We wish to thank G. B{\' e}langer for a very valuable help in this work.

\label{lastpage}
\end{document}